# Amorphous-to-crystalline transformation: how cluster aggregation drives the multi-step nucleation of ZIF-8

Ahmet R. Dok ‡, Sambhu Radhakrishnan§,¶, Flip de Jong‡‡, Estelle Becquevort§,¶, Olivier Deschaume†, Vinod Chandran C. §,¶, Yovan de Coene††, Carmen Bartic†, Mark Van der Auweraer‡‡, Wim Thielemans§§, Christine Kirschhock¶, Monique A. van der Veen¶¶, Thierry Verbiest††, Eric Breynaert*§,¶, Stijn Van Cleuvenbergen*‡

‡Department of Chemistry, Molecular Imaging and Photonics, KULAK – KU Leuven, E. Sabbelaan 53, 8500 Kortrijk, Belgium

§NMR-Xray Platform for Convergence Research (NMRCoRe), KU Leuven, 3001 Leuven, Belgium

‡‡Department of Chemistry, Molecular Imaging and Photonics, KU Leuven, Leuven Chem&Tech, Celestijnenlaan 200F, 3001 Leuven, Belgium

†Department of Physics and Astronomy, Soft Matter Physics and Biophysics Section, KU Leuven, Celestijnenlaan 200 D - box 2416, 3001 Leuven, Belgium

††Department of Chemistry, Molecular Imaging and Photonics, KU Leuven Celestijnenlaan 200D, 3001 Heverlee, Belgium

§§Department of Chemical Engineering, Sustainable Materials Lab, KU Leuven, campus Kulak Kortrijk, Etienne Sabbelaan 53, 8500 Kortrijk, Belgium

¶Center for Surface Chemistry and Catalysis - Characterisation and Application Team (COK-KAT), KU Leuven, 3001 Leuven, Belgium

¶¶Catalysis Engineering, Department of Chemical Engineering, Delft University of Technology, 2629 Delft, The Netherlands

**ABSTRACT:** Nucleation, the pivotal first step of crystallization, governs essential characteristics of crystallization products, including size distribution, morphology, and polymorphism. While understanding this process is paramount to designing industrial production processes, major knowledge gaps remain, especially for porous solids. For nanocrystalline ZIF-8, one of the most widely studied metal-organic frameworks, questions regarding the species involved in nucleation and their structural and chemical transformations remain unanswered. By combining harmonic light scattering, inherently sensitive to structural changes, with NMR spectroscopy, revealing molecular exchanges between particles and solution, we captured the crystallization mechanism of ZIF-8 in unprecedented detail. Upon mixing, small charged prenucleation clusters (PNCs) form, exhibiting an excess of ligands and a net positive charge. We show that nucleation is initiated by aggregation of PNCs, through the release of ligands and associated protons to the liquid. This leads to the formation of charge-neutral amorphous precursor particles (APPs) which incorporate neutral monomers from the solution and crystallize into ZIF-8. Our work highlights chemical dynamics as a vital, yet often overlooked, dimension in the multi-stage structural evolution of MOFs. By establishing the critical role of PNCs in the nucleation of ZIF-8, new pathways open up for controlling crystallization of metal-organic frameworks through targeted chemical interactions.

## INTRODUCTION

Metal-organic frameworks (MOFs) are a class of crystalline materials that form nanoporous networks by interconnection of metal-based nodes with polydentate organic ligands. Owing to their large surface area and intrinsic structural and functional diversity, MOFs can be tuned for specific applications including gas storage, separation, and catalysis. This has led to a tireless exploration of the MOF material space, with nearly a hundred thousand structures reported so far.[1] The concept of secondary building units (SBUs) has served as a basis to guide this endeavor.[2] However, despite the success of modular chemistry in rationalizing MOF synthesis, there is a gap between the selection of appropriate building blocks and the final crystal product.[3] To date, the discovery and optimization of MOFs have relied heavily on trial and error. In order to bridge this gap, a fundamental understanding of their crystallization process is essential.[3] The nucleation and growth of porous crystals, including MOFs and zeolites, has become an active research field in recent years. Insights into the thermodynamics, phase transformations, and kinetics have been obtained for a number of prominent materials, including relevant energetic barriers and (nonclassical) nucleation and growth mechanisms.[3] Crystallization processes of MOFs typically follow a structured progression involving multiple steps before reaching the final crystalline product, the specific timeline of which can vary significantly depending on the experimental conditions.[4] To effectively optimize and manage these crystallization processes, it is crucial to gain a comprehensive understanding of the sequential events that occur rather than solely focusing on the characterization of the final crystalline product. To accomplish this objective, employing experimental approaches that enable multidiagnostic *in situ* monitoring of crystallization processes is imperative.

ZIF-8, one of the most widely investigated MOFs due to its excellent gas separation abilities and high stability, exemplifies the current challenges encountered in managing crystallization processes. For large-scale production of ZIF-8, fast room-temperature protocols have been developed aimed at the production of nanocrystals in a straightforward and reproducible manner. However, the upscaling of these synthesis routes is currently hampered by the incomplete knowledge of the crystallization mechanism.[5] Despite the straightforward synthesis of ZIF-8 by mixing an excess of 2-methylimidazole (2-MeIm) and $Zn(NO_3)_2 \cdot 6H_2O$ in a methanolic solution,[6] understanding the initial stages of its crystallization mechanism has proved remarkably complex, as most of the crystal growth takes place within minutes.[7] Detailed insights into the early stages were obtained through *in situ* synchrotron small- and wide- angle X-ray scattering (SAXS/ WAXS).[8–10] SAXS data showed the formation of a population of prenucleation clusters (PNC) with a stable size of about 2 nm almost immediately after mixing. These PNCs were consumed during particle growth, yet it remains uncertain whether they simply serve as a monomer reservoir or actively participate in particle nucleation.[6,8,9] The initial structure of the growing particles remained an open question as well, since WAXS detection of the crystalline framework was slightly delayed with respect to the initial detection of particle formation via SAXS. The authors speculated this might indicate the involvement of amorphous precursor particles (APP) that subsequently transform into crystalline ZIF-8.[8] Unfortunately, the time between the onset of the SAXS and WAXS signals was too small with respect to the experimental time resolution to formulate a definitive conclusion.[6,8,11] The notion of the potential importance of APP is however strengthened by other studies, showing the presence of amorphous domains coexisting within ZIF-8 crystallites through ex-situ measurements, or establishing the disordered nature of the nanosized PNCs preceding nucleation of ZIF-8.[10,12–18] The involvement of APP has moreover been invoked in aqueous conditions or for ZIF frameworks with different topologies.[4,19–21] For the crystallization of structurally related zeolites, analogous two-step nucleation mechanisms involving APP or ion-paired PNCs have also been observed.[22–26] Finally, the structural development of ZIF-8 cannot be separated from the chemistry occurring in the mother liquor of the crystals. In the case of ZIF-8, a series of ligation and deprotonation reactions results in a complex chemical landscape dominated by monomeric zinc complexes almost immediately after mixing.[10,27] The pathways connecting this diverse pool of dissolved species to the final crystal structure remains a matter of debate.[6,10,27,28] Given that initial complexes and oligomers are overstoichiometric compared to the final composition of ZIF-8, $Zn(MeIm)_2$, the nucleation and growth process involves chemical reactions that significantly impact the kinetics of the different stages.[29]

In this study, a two-tiered approach combining harmonic light scattering (HLS) experiments and static nuclear magnetic resonance (NMR) spectroscopy was adopted to address the questions surrounding the structural and chemical evolution during the crystallization process of ZIF-8. HLS, like other nonlinear optical methods, is particularly sensitive to structure and shows promise for studying crystallization processes.[30–38] For second harmonic scattering (SHS), this structural sensitivity is illustrated most clearly by the effect of an inversion center, which results in an effective cancellation of the response within the dipole approximation. In other words, SHS detects only non-centrosymmetric structures. In an earlier study, SHS was used to monitor the non-centrosymmetric I-43m space group of ZIF-8 during crystallization.[30] To probe the potential involvement of APPs in the crystallization pathway, we introduced third harmonic scattering (THS) as an additional probe in this study. In contrast to SHS, THS can identify all symmetries, making it possible to differentiate between amorphous phases (exclusively via THS) and crystalline ZIF-8 (via both SHS and THS).[39] Through polarization measurements, additional information about (point group) symmetry can also be derived.[30,39,40] Where HLS measures the structure of the emerging solid phase, static NMR spectroscopy was employed to deliver the inverse picture by selectively probing chemical speciation in the mother liquid. Static NMR is regularly used in crystallization studies to track solvent dynamics, solute interactions, and the impact of temperature and concentration on crystallization.[41–43] In this work, static NMR spectroscopy was used to monitor the evolution of the concentration and speciation of dissolved ZIF-8 linkers and small oligomers, as well as the concentration of mobile protons. By combining both techniques, a complete picture of all events occurring in the solid and liquid fractions of the system could be obtained, capturing the different stages of ZIF-8 crystallization. We show that small positively charged PNC form neutral APP by condensation and aggregation. Subsequently APP transform into crystalline ZIF-8 by local nucleation and growth processes. As crystallization progresses and the PNC reservoir becomes depleted, the APP are consumed and the growth mechanism switches to solution-mediated growth, in line with Ostwald ripening.

## RESULTS

**Nanoparticle formation detected by light scattering.** The fast crystallization of ZIF-8 in methanol was measured in situ with a range of time-resolved scattering techniques for a molar ratio of 1:4:1000 (Zn: 2-MeIm:MeOH). Static light scattering (SLS), and HLS were conducted simultaneously using a continuous 543 nm laser source and a pulsed laser source at 1260 nm, respectively, for the same synthesis. While dynamic light scattering (DLS) measurements were performed separately under the same conditions. The synthesis protocol is highly reproducible when starting from the same reagent and solvent batches. Between solvent batches, small variations in induction times were observed, but overall the data shows the same trends. The small observed variations are attributed to slight differences in water content (see ESI.2.8). Due to the inherent confocal effect of SHS and THS,[40] SLS probes a much larger volume than SHS and THS and, consequently it is more sensitive to multiple scattering effect (See Fig. ESI.2.6). To limit the impact of multiple scattering effects (due to particle aggregation and precipitation) on the data analysis, either through depolarization or intensity fluctuations, the analysis was focused on the first 500 seconds of the synthesis. Both DLS and SLS indicated multiple scattering only became significant after this timeframe (See ESI.1a and ESI.2.6). Immediately after mixing, a stable scattering signal was detected for SLS and SHS, but not for THS, as shown in Fig. 1a. These signals can be attributed to the molecular (hyper-)Rayleigh scattering stemming mainly from the aromatic 2-MeIm linkers and the solvent molecules. After an induction period of 52s, labeled $\tau_p$, SLS is the first technique that detects particle growth showing a sharp increase in scattering intensity. THS and SHS register a rapid increase in signal after approximately 72 and 84 seconds. Since, as laid out in more detail below, the increase in SHS is in line with the formation of ZIF-8, the onset time at 84 seconds is labeled $\tau_{ZIF-8}$. It is noteworthy that after the steep initial increase, reminiscent of burst nucleation, the curves do not plateau in the typical sigmoidal manner expected for classical nucleation and growth. Indeed, while the process decelerates after approximately 200 seconds, all scattering signals still display a marked increase in intensity, by a factor of 2.4, 2.7, and 2.3 for SLS, SHS, and THS respectively, between 200 and 500 seconds.

Time-resolved DLS measurements, performed under identical synthesis conditions and using the finest possible time-resolution to yield a reasonable signal-to-noise ratio, allowed tracking the evolution of particle size. Figure 1b shows the evolution of the growing nanoparticles with time.

The first nanoparticles were detected around 50 seconds. At 70 seconds, a diameter around 40 nm was determined with good accuracy. Thereafter, the effective particle diameter continuously increased until a size of about 65 nm was reached at around 180 seconds. After that point, the particle size increased at a slower rate reaching 80 nm at 500 s. These results are in line with SLS measurements by Cravillon et al., who observed only minor changes in the overall size of ZIF-8 nanocrystals after 130 seconds under the same conditions.[6] The observed stages could be attributed to a nucleation and growth process followed by Ostwald ripening (OR) by fitting to a combined Johnson–Mehl–Avrami–Kolmogorov (JMAK) and Lifshitz-Slyozov model:[44]

$$V_{(t)} = (\{1 - exp[-(k_g t)^n]\} + (t - \tau_{OR})k_{OR})V_{lim} \qquad (1)$$

The first term in equation 1 follows from the Johnson–Mehl–Avrami–Kolmogorov (JMAK) equation, with $k_g$ a growth rate constant and n the Avrami exponent. The second term describes Ostwald ripening, with $k_{OR}$ the rate constant and $\tau_{OR}$ the onset time of Ostwald ripening as shown in table 1.

**Table 1. Fitted parameters of the JMAK-Lifshitz-Slyozov model to the measured DLS data.**

| JMAK-LS parameters | Fitted results | Description |
| --- | --- | --- |
| $k_g$ | $0.43 \pm 0.047$ min$^{-1}$ | JMAK growth rate constant |
| n | $3.5 \pm 0.82$ | Avrami exponent |
| $k_{OR}$ | $0.14 \pm 0.03$ min$^{-1}$ | Rate constant OR |
| $\tau_{OR}$ | $180 \pm 69$ s | Onset time OR |

The Avrami coefficient obtained here through DLS is significantly higher than the value determined by WAXS measurements for the same synthesis.[3,45] While WAXS selectively detects the crystalline phase, DLS detects light scattering by particles, irrespective of their crystalline or amorphous nature. This mismatch may therefore point to the presence of phases without long-range periodic order. The Avrami exponent of 3.5 is in line with three dimensional growth, although the interpretation of the Avrami exponent with respect to the crystallization mechanism is generally considered problematic for solution crystallization processes.[45] Also note that the JMAK model does not specify the nucleation mechanism, which could be classical or non-classical.[44]

Unlike conventional light scattering techniques, HLS provides a measure of the symmetry of the scattering species during the crystallization process through measurement of the depolarization ratio $\rho$. Figure 1c shows the evolution of $\rho_{SHS}$ and $\rho_{THS}$. Immediately after mixing, $\rho_{SHS}$ has a value of ~0.3. This value reflects the symmetry of the 2-MeIm linkers,[30] demonstrating that these species initially dominate the SHS response. When the SHS signal starts to increase rapidly around 84 seconds at $\tau_{ZIF-8}$, $\rho_{SHS}$ gradually shifts to a value of approximately 0.63 in line with the expected theoretical value of 2/3 for the $T_d$ point group of ZIF-8. This confirms the formation of crystalline ZIF-8, which was further validated through ex situ powder diffraction of the final product (ESI.4.1). For THS, which is typically weaker, no measurable signal was detected prior to the onset of particle formation. $\rho_{THS}$ assumes a value close to 0 throughout the entire process. As laid out in the materials and methods, such low values for $\rho_{THS}$ are consistent with the formation of the highly symmetrical structure of ZIF-8 as well as APPs.

Since SHS serves as an exclusive probe for crystalline ZIF-8, while THS probes ZIF-8 as well as APPs, comparison of both responses can reveal potential contributions of an amorphous precursor. We introduce the THS to SHS ratio for this purpose. Based on equations M3 and M4 laid out in the materials and methods, $THS/SHS$ can be reduced to:

$$\frac{THS}{SHS} \sim \frac{\sum_{zif-8,i} N_i <|\chi^{(3)}_{HRS,i}|^2><V_i^2> + \sum_{a,j} N_j <|\chi^{(3)}_{HRS,j}|^2><V_j^2>}{\sum_{zif-8,i} N_i <|\chi^{(2)}_{HRS,i}|^2><V_i^2>} \qquad (2)$$

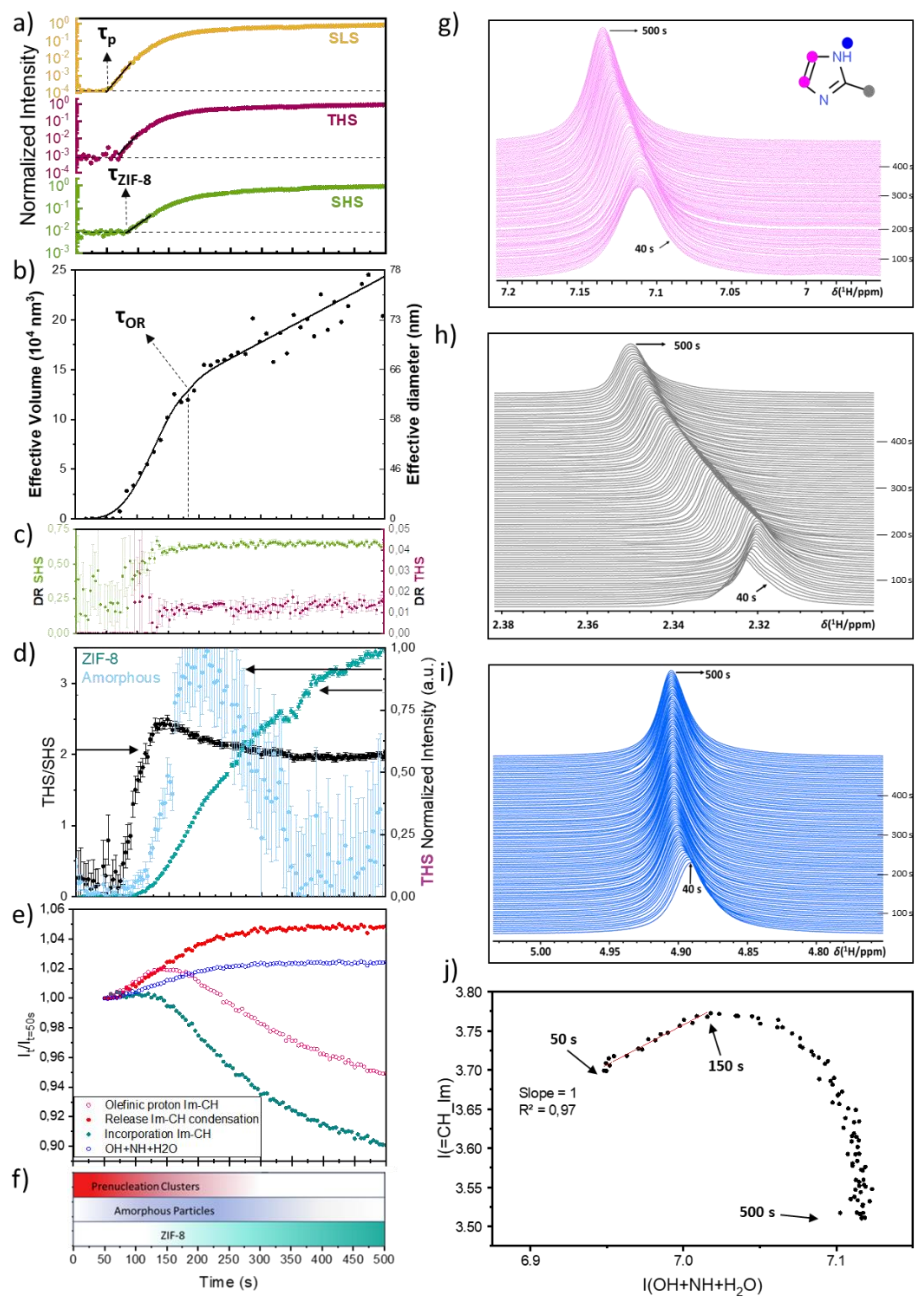

Figure 1 Time-resolved, in situ light scattering and NMR spectroscopy data collected during the crystallization of ZIF-8 from a 1:4 molar mixture of Zn(NO$_3$)$_2$ and 2-methylimidazole. **a.** Normalized total scattering intensity (VH+VV scattering) for SLS, SHS and THS. The onset of the scattering for all harmonics was determined by intersecting a linear fit of the initial onset (black line) with the mean scattering intensity before onset (dashed line). Details are provided in ESI.2.5. **b.** Evolution of effective diameter and averaged particle volume obtained by DLS (points) and a Lifshitz and Slyozov model fit (black line) describing the data as a combination of sigmoidal Avrami growth curve, followed by Ostwald ripening. **c.** Time-resolved depolarization ratio for SHS (green) and THS (purple). **d.** Black curve: Time-evolution of the THS/SHS ratio for ZIF-8 crystal growth (left y-axis). THS and SHS signals onset around 70 and 80s respectively. Blue and green curves: THS scattering intensity evolution of the amorphous and crystalline fractions (right y-axis). **e.** Time-evolution of the integrated area of $^1$H NMR resonances normalized to itself at T=50s, associated to the olefinic 2-methyl imidazole resonance (pink, open spheres) and , exchangeable protons corresponding to OH + NH+ H$_2$O (blue open spheres); The pink trace of the olefinic resonance is theoretically deconvoluted into two traces corresponding to ligand release via condensation (red closed spheres) and ligand assimilation into the ZIF-8 crystals (cyan closed spheres). **f.** Distribution of the prenucleation clusters, amorphous precursor particles and ZIF-8 particles throughout the 4 stages of the crystallization reaction. **g.** Time evolution of the $^1$H NMR resonances of the olefinic protons of 2-methylimidazole **h.** Time evolution of the $^1$H NMR resonance of the methyl resonance of 2-methylimidazole. **i.** Time evolution of the composite resonance of exchangeable protons (NH, OH and H$_2$O). **j.** Correlation plot of the absolute integrals Figure ESI.3.1) of the olefinic $^1$H resonance of 2-methyl imidazole plotted (y-axis) and the composite resonance of exchangeable protons (x-axis). The synthesis time evolves from left to right. The red line indicates a linear fit in the timeframe between 50 and 150 s.

In equation 2, we omitted the negligible molecular contribution of the solvent and the solute linker molecules. Subscripts *a* and *zif-8* refer to amorphous and crystalline particles (or domains) respectively. Remark that the $THS/SHS$ signal is proportional to the intensity of the laser source, but since the intensity remains constant during the measurement $I_\omega$ was omitted. Before discussing the results, it is instructive to consider $THS/SHS$ within the framework of a classical nucleation process, characterized by the presence of a single crystalline species throughout the entire process. Under such circumstances, $THS/SHS$ can be reduced to:

$$\left(\frac{THS}{SHS}\right)_{CNT} \sim \frac{N_{zif-8} <|\chi^{(3)}_{HRS,zif-8}|^2 > \langle V^2_{zif-8}\rangle}{N_{zif-8} <|\chi^{(2)}_{HRS,zif-8}|^2 > \langle V^2_{zif-8}\rangle} = constant \quad (3)$$

Due to the proportional increase of both the volume and number of particles, a constant value for $THS/SHS$ is thus expected in the case of a classical nucleation scenario (CNT). Stated otherwise, deviations from a constant $THS/SHS$ point to additional contributions from amorphous intermediates. Contribution from defect formation in the ZIF-8 crystal lattice is not expected, as no deviation from perfect stoichiometry is observed through *ex situ* NMR measurements (ESI.4.2 NMR analysis).

For the synthesis of ZIF-8, $THS/SHS$ is plotted in Figure 1d. It is clear that $THS/SHS$ varies significantly over the course of the process, and hence the crystallization cannot be described according to a classical nucleation model. In the initial stage of nucleation and growth, $THS/SHS$ increases sharply, which can only be attributed to additional contributions to the THS signal related to the formation of APPs. Even after the start of the formation of crystalline ZIF-8 at $\tau_{ZIF-8}$, $THS/SHS$ continues to increase. This implies that the amorphous phase develops at a more rapid pace compared to the formation of crystalline ZIF-8. A plot of the growth rate for SHS and THS, calculated as the first derivative of both signals, further corroborates this (Fig. ESI.2.9). Around 150 seconds, $THS/SHS$ reaches a maximum, after which a gradual decrease ensues. After 350 seconds, $THS/SHS$ plateaus, as SHS and THS increase proportionally from that point onwards. In other words, after 350 seconds, the amorphous to crystalline transition is complete and SHS and THS exclusively detect the crystalline structure of ZIF-8. We can now write:

$$\left.\frac{THS}{SHS}\right|_{t>350\,s} = \frac{THS_{zif-8}}{SHS_{zif-8}} = 1.94 \quad (4)$$

Through substitution we can now express the total THS intensity as:

$$THS_{tot} = THS_a + THS_{zif-8} = THS_a + 1.94 \times SHS_{zif-8} \quad (5)$$

Since $SHS_{zif-8}$ corresponds to the total intensity detected by SHS ($SHS_{tot}$), it becomes possible to separate the contributions of the amorphous and crystalline fraction throughout the process, as plotted in Figure 1d. This paints a comprehensive picture of the crystallization process. The crystallization of nanocrystalline ZIF-8 starts with the formation of APPs which amplify throughout the initial stage. Shortly after, crystalline ZIF-8 starts forming at a slower pace. Around 180 seconds at $\tau_{OR}$, the amorphous fraction reaches a maximum. This coincides with the switch in growth mechanism to Ostwald ripening detected by DLS (Fig. 1b). Further transformation of the APPs ensues until it is completely consumed around 350 seconds. Comparing the THS intensities for the amorphous ($THS_a$) and the crystalline fraction ($THS_{zif-8}$), $THS_a$ is notably lower than $THS_{zif-8}$ throughout the process. This might be related to the relative size of amorphous domains, as THS scales with the volume squared, and to their density, through $\chi^{(3)}_{HRS,a}$.

**Solution chemistry measured by static ¹H NMR.** For the *in situ* ¹H NMR investigation into the formation of ZIF-8, solutions of Zn(NO$_3$)$_2$ and 2-methylimidazole in methanol-d$_4$ were prepared. Pure 2-MeIm dissolved in methanol-d$_4$ exhibits two distinctive ¹H resonances at chemical shifts of 2.3 ppm and 6.8 ppm corresponding to the methyl and olefinic protons, as depicted in Figure ESI.3.1. The FHWM of these resonances were 2.5 Hz and 2 Hz respectively. Apart from the imidazole resonances, resonances corresponding to methyl groups of residual methanol solvent at 3.3 ppm and the composite resonance of NH of imidazole and OH of methanol at 5 ppm were observed. For *in situ* static ¹H NMR experiments, the two solutions were mixed in a 1:4 molar ratio in a 5 mm NMR tube at time 0 s and immediately inserted in the NMR probe head, and spectra were acquired every 5 s and the results are depicted in Figures 1e-j. At 40 s, a first qualitative spectrum was acquired and from 50 s onwards, the field homogeneity and stability of the system was suitable to allow recording quantitative spectra (Figure 1g-i, Figure ESI.3.1). This is demonstrated for the trace methyl protons still present in the methanol-d4 solvent. The integral of the resonance associated with non-exchangeable methyl protons of methanol is indeed constant as shown by the red trace in Figure ESI.3.2b. This signal therefore provides an internal reference for relative quantification of all other resonances associated with dissolved species, larger than methanol (i.e. t1 ≅ t2; t1(species) > t1(methanol)). The use of static ¹H NMR implies that all observed resonances pertain exclusively to mobile protons, implying the signal exclusively reflects protons present on dissolved components and small nanoaggregates: i.e. dissolved Zn$^{2+}$-2-MeIm complexes, inner- or outer-sphere nanoaggregates of 2-MeIm complexes, free (unreacted) 2-MeIm, and exchangeable protons. Upon introduction of Zn(NO$_3$)$_2$, the olefinic resonance of 2-MeIm underwent substantial alterations, broadening to a FWHM of 40 Hz and shifting to 7.1 ppm (Figure ESI.3.1). The transition to a higher chemical shift is attributed to the complexation of Zn$^{2+}$ ions by 2-methylimidazole. The pronounced broadening results from chemical shift broadening effects, arising from the formation of a score of Zn$^{2+}$-2-MeIm$^-$ and Zn$^{2+}$-2-MeImH complexes, and from exchange phenomena occurring between various molecular species in the solution (i.e. 2-MeImH, 2-MeImH$_2^+$, and 2-MeIm$^-$ incorporated in Zn$^{2+}$-complexes), nanoaggregates and exchangeable surface species of nanoparticles.

Detailed examination of the evolution of the resonances in the ¹H spectra revealed concurrent variations in the chemical shifts and integrated area (Figure 1e,j; Figures ESI.3.1-2). The non-exchangeable methyl proton resonance associated with the methanol solvent was used as internal reference for relative quantification (infra). The integrated areas corresponding to the non-exchangeable imidazole olefinic ¹H resonances displayed an initial increase in the timeframe between 50 and 100 s, after which the signal plateaued, before starting to gradually decline again at 150 s (Figure 1e). This initial increase is indicative of an increase in concentration of mobile (dissolved) 2-MeIm molecules. It indicates a release of 2-MeIm molecules from an immobile (solid) phase into the liquid phase. Concurrently, the area of the $^1$H resonance associated with the exchangeable protons in the system - the NH, $H_2O$ and OH protons (Figure 1e) - steadily increased as function of time up to around 200 s after mixing after which it plateaued. This suggests a release of exchangeable $^1$H species from clusters containing immobile and thus invisible 2-MeIm molecules. Linear fitting of the increase in the =CH area versus the area of the resonance associated with exchangeable protons reveals a slope of 1, hinting at release of two exchangeable protons per imidazole molecule released (Figure 1j). This is in line with particle growth by condensation via release of $MeImH_2^+$, as proposed by Yeung et al.[29] Owing to rapid protonation/deprotonation reactions, the charged $MeImH_2^+$ species quickly equilibrate with all dissolved species in solution, resulting in the overall release of dissolved linker molecules (MeImH) and protons in a ratio of 1:1. The subsequent decrease after 150 s suggests continuous assimilation of ligand into a solid phase. The 1:1 correlation between the release of exchangeable protons and linkers into solution upon condensation invites differentiation of the =CH signal into separate traces representing condensation and incorporation of neutral linkers (Figure 1e). Initially, condensation dominates linker release, but around 150 seconds, incorporation from solution becomes prevalent, with some overlap. Since the resonance from linker incorporation does not correlate with the exchangeable protons in solution, it can be inferred that the incorporated species are neutral, , e.g. MeImH or $Zn(MeIm)_2$. An initial attachment of positively charged species, with rapid release of extra linker and charge into the liquid, however, cannot be excluded.

The observed changes in the $^1$H chemical shifts also provide a similar picture (Figure 1g-i). The olefinic resonance initially underwent a subtle shift to lower chemical shift, primarily up to 100 s after mixing, followed by a progressive transition to higher chemical shift (Figure 1f). The initial effect is attributed to the increased presence of free 2-MeIm molecules, indirectly indicating a depletion of Zn-ions. The transition after 100 s aligns with the crystallization of the solid fractions, indicative of intensified interactions of the 2-MeIm moieties with $Zn^{2+}$ species and the solid constituents. The composite resonance corresponding to the OH+NH+$H_2O$ resonances initially also showed an evolution towards higher chemical shift, consistent with the release of 2-MeImH$_2^+$(Figure 1i). Note that the resonances associated with methyl protons on 2-MeIm were explicitly excluded from the above analysis. In the context of an evolving mixed-phase system, it is important to consider the potential impact of the quantum rotor effect of methyl groups residing in nanoparticles. In static NMR, resonances associated with immobile species (i.e. species that are part of the nanoparticles) typically broaden beyond detection due to dipolar interactions. In magic angle spinning NMR (MAS NMR), this broadening is counteracted by spinning the sample at the magic angle. The quantum rotor effect of methyl groups can however have a similar effect as MAS, especially for species that are part of small nanoparticles. This can result in enhanced visibility of protons associated with methyl groups in the nanoparticles as compared to types of protons in the same particles. As this can impact the quantification of mobile versus immobile species in the system, the methyl resonance of imidazole was excluded from quantitative analysis.

Whereas light and X-ray scattering experiments exclusively detect bigger clusters and solid particles, i*n situ* static $^1$H NMR spectroscopy provides information about mobile components in a system, and their evolution in time. Static $^1$H NMR thus provides complementary information and assists in obtaining a comprehensive picture of the crystallization process. The $^1$H NMR spectroscopy data indicates formation of oligomeric species upon mixing, which are not detectable in the static NMR spectrum. It also points towards the release of 2-MeImH and $H^+$ (ratio1:1) from these oligomers during the initial nanoparticle growth and suggest that the predominant imidazole species assimilated in the crystalline ZIF-8 formation phase is 2-MeIm$^-$, potentially as a complex including $Zn^{2+}$ cations. The timescales of these findings are in alignment with those observed in the light scattering experiments. Finally, in a separate *ex situ* experiment, the quantity of linker in the solid ZIF-8 was found through quantitative NMR to closely match the expected perfect stoichiometry, indicating a defect-free final product.[46,47]

## DISCUSSION

Based on our results, we identified four key stages in the nucleation and growth mechanism of nanocrystalline ZIF-8 in methanol, which will be discussed in detail below. The onset of these stages ($\tau_p$, $\tau_{ZIF-8}$, $\tau_{OR}$) are indicated in Fig. 1a and 1b. Immediately after mixing, complexation leads to the formation of nanosized clusters, as detected by NMR. At $\tau_p$ (50 s), particle growth is first detected by SLS. At $\tau_{ZIF-8}$ (84 s), the appearance of crystalline ZIF-8 is confirmed through SHS measurements. Finally, at $\tau_{OR}$ (180s), the onset of Ostwald ripening is detected by a fit to the DLS data.

Prior to the detection of the first nanoparticles at $\tau_p$, static $^1$H NMR spectroscopy is highly sensitive to dissolved species and mobile species present in small nanoaggregates. However, as it takes some time to achieve a $B^0$ field homogeneity, quantitative results were only acquired from $\tau_p$ onwards. Static $^1$H NMR measurements nevertheless indirectly provided information about the chemical species formed prior to $\tau_p$. Contrary to expectations, the detection of the first particles coincided with an increase in the concentration of mobile imidazole in the solution (Fig. 1e). Since NMR exclusively detects species in free exchange with the solution and not those that are in the bulk of nano-particles, this implies that the imidazole molecules were initially immobilized in extended structures, and subsequently released into solution. Different studies have demonstrated that a diverse array of complexes, varying in coordination geometry and the number of coordinating imidazole molecules, forms almost immediately after mixing.[10,27,45,48] In the case of ZIF-8 synthesis in methanol, using a 1:4 metal to linker ratio, ESI-MS identified monomeric tetrahedral complexes with three methylimidazole ligands as the predominant species, aligning with similar findings in Co ZIF-67.[27] Alongside these small complexes, the formation of non-crystalline clusters preceding particle growth, a few nanometers in size, has been detected by synchrotron based experiments.[7,10,19] Since static liquid state $^1$H NMR is able to detect ligands present in monomeric complexes but not in clusters, it follows that clusters act as the source of imidazole and proton release detected after $\tau_p$. The continued release of linker from clusters during particle growth moreover indicates that they play a direct role in particle growth – not merely serving as nucleation sites followed by monomer attachment, as this would rapidly lower the concentration of mobile imidazole in the solution. Through correlation of the area of the olefinic imidazole resonance with that of the exchangeable protons we were able to assign the released species as 2-MeImH and $H^+$ in a 1:1 ratio (Figure 1j). This aligns well with a kinetic model presented by Yeung et al., who predicted that rapid equilibration of monomeric complexes leads to the formation of oligomeric prenucleation clusters, or PNCs, in a pre-equilibrium state.[45] Since these PNCs contain excess protonated ligand as compared to the final ZIF-8 stoichiometry, $Zn(MeIm)_2$, further growth involves the release of superfluous linkers and protons through condensation. The increase in mobile imidazole detected by NMR after the onset of particle growth at $\tau_p$, is therefore directly related to condensation reactions from within and between PNCs. This implies that particle growth is driven by the aggregation of PNCs rather than incorporation of monomeric species or cluster dissolution.

After $\tau_p$, SLS and DLS reveal an initial exponential increase in particle size, that fits to the JMAK model (Fig. 1a and 1b). Subsequently, at $\tau_{ZIF-8}$, the crystalline phase of ZIF-8 was detected by SHS with a 35 second delay. The structural assignment of ZIF-8 was further corroborated by the SHS depolarization ratio, aligning with the I-43m space group of the crystal lattice (Fig. 1c). This observed delay in detection is reminiscent of the findings of Cravillon et al., who reported a similar lag in detection of ZIF-8's crystalline phase using synchrotron WAXS in comparison with SAXS. Since it was not clear whether this difference could be attributed to the presence of APPs, or merely to differences in sensitivity between both techniques, the authors were not able to draw definitive conclusions. In this study, SHS, akin to WAXS, demonstrated exclusive sensitivity to the crystalline form of ZIF-8, but exhibits limited sensitivity compared to SLS and DLS. Through theoretical and experimental characterization of the SHS efficiency of crystalline ZIF-8, we estimated a minimum detectable diameter of around 10 nm, which is significantly smaller than the particle size of 40 nm detected by DLS at $\tau_{ZIF-8}$ (see ESI.6). This implies that the particles detected by DLS are not entirely crystalline at this point. By incorporating THS, which, unlike SHS, is sensitive to amorphous phases, we were able to unambiguously confirm the presence of APPs. Their amorphous nature was corroborated through depolarization measurements (Fig. 1c). Through analysis of the relative progression of SHS and THS, we were moreover able to separate the evolution of the amorphous and crystalline fractions over time (Fig. 1d). Our analysis demonstrates that the formation of APPs precedes crystalline ZIF-8 formation. These APPs do not merely act as short-lived intermediates but rather drive the nucleation and growth process throughout the exponential growth stage. Growth rate analysis indeed reveals that the formation of APPs consistently outpaces that of the crystalline phase during this stage (see Figure ESI.2.9). As detailed above, NMR data indicate an increase of ligand concentration in the liquid phase upon initial formation of the APPs, consistent with PNC association, even before the appearance of crystalline ZIF-8. This implies that APPs are formed through association of PNC and subsequently transition into crystalline ZIF-8. The continuing release of linker and protons into the solution, even after the onset of crystallization, suggests that nucleation and growth of the crystalline framework occurs through condensation and particle reorganization rather than through bulk dissolution of the APPs and monomer attachment to isolated nuclei.

At $\tau_{OR}$ (180 seconds), the growth mechanism shifts from exponential growth to diffusion controlled Ostwald ripening. This shift is evidenced by a Lifshitz-Slyozov fit of the DLS data (Fig. 1b), and is marked by a reversal of the flux of mobile ligands observed by NMR (Fig. 1e), which now shows a decrease. Concurrently, the rise in proton concentration registered by NMR starts to level off. This implies that further growth predominantly occurs through incorporation of monomeric species with a net stoichiometry of the product through Ostwald ripening, rather than through association-condensation. The amorphous fraction, as detected by THS (Fig. 1d), reaches a maximum at $\tau_{OR}$, after which it starts to decrease. Synchrotron-based X-ray measurements showed a further increase in crystallinity throughout this stage,[7,49] suggesting reorganization, either intraparticle or most likely by local dissolution re-precipitation on the particle surface. Around 350 seconds, the amorphous fraction is fully depleted, while the proton concentration registered by NMR reaches a constant value. At this time, the reservoir of PNC building blocks has fully transformed into crystalline ZIF-8. Beyond this point, growth proceeds exclusively through the incremental addition of monomers.

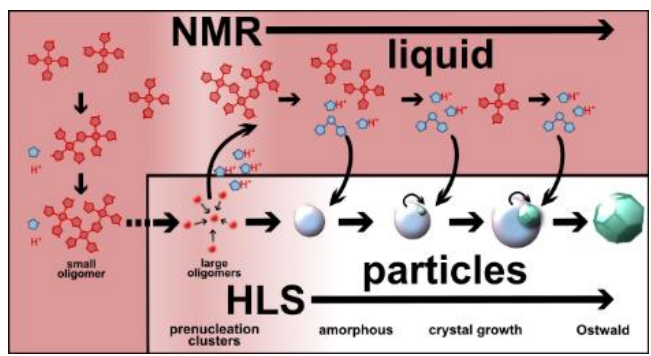

Figure 2 Scheme showing the multi-step nonclassical nucleation mechanism as derived from our observations through NMR and HLS. The key stages involved in the mechanism shown are the formation of non-crystalline PNCs through oligomerization followed by the aggregation of the PNCs into APPs by condensation. The further formation of crystalline ZIF-8 is driven by particle reorganization and condensation followed by diffusion controlled Ostwald ripening. The positively charged species are indicated in red, while neutral species are shown in blue. The crystalline framework is shown in green.

Our observations are in line with a three-step nucleation mechanism followed by Ostwald ripening, as summarized in Fig. 2. Immediately after mixing, a population of PNC, with a stable size of a few nanometers, forms alongside with a variety of Zn-linker complexes, mainly consisting of monomeric species. These PNC are positively charged and contain an excess of ligand and protons compared to the final structure of ZIF-8. Amorphous precursor particles grow through association of PNCs. The growth of APPs is accompanied by the release of excess ligand and protons into the surrounding solution, resulting in the formation of larger, net neutral charge particles. Once these particles become neutral, they can incorporate neutral species from the solution. Thus, at some point, the aggregation of PNCs and the addition of units from the liquid occur in parallel—initially dominated by massive aggregation, with monomer addition taking over in later stages. Nucleation of crystalline ZIF-8 then takes place inside these APPs, after which crystal growth ensues through particle reorganization, most likely through local dissolution-reprecipitation. When the reservoir of PNCs becomes depleted, the growth mechanism switches to Ostwald ripening, and further growth occurs through incorporation of charge neutral monomeric species from solution.

The different steps of the nonclassical pathway can be understood from their progression in terms of charge and stoichiometry, transitioning from positively charged, over-stoichiometric species to the final stoichiometric composition of the neutral charge crystalline phase. An important aspect not directly observed in our study, but inherently present and influencing the system, involves the positively charged prenucleation species and their counter ions, notably nitrate ions ($NO_3^-$), which likely serve as charge compensating species. Although our methods do not detect these ions, their presence and role in balancing the positive charge of the clusters are expected to significantly impact the nucleation mechanism. A study by Priandani et al. on the modulation of $Zn(NO_3)_2$ based ZIF-8 synthesis with sodium chloride supports this notion, highlighting how varying amounts of NaCl can lead to distinct changes in crystal size and morphology, akin to those observed in solvothermal synthesis with $ZnCl_2$.[50] This opens up the exciting possibility of rational crystallization control through interaction with intermediate species in the nucleation pathway.

## CONCLUSION

In conclusion, our research provides unprecedented insights into the crystallization process of ZIF-8, delineating a three-step nucleation mechanism driven by chemical dynamics. This model progresses from the formation of positively charged PNC to APPs, leading to nucleation of the crystalline phase and culminating in crystal growth via particle reorganization and Ostwald ripening. Our findings emphasize the critical role of integrating in situ chemical and structural analysis to unravel the complex interplay between structure and chemistry that drives crystallization of MOFs. By detecting the chemical nature of the involved species, our approach not only deepens our understanding of the crystalline phase formation of ZIF-8 but also provides handles for controlling crystallization in MOFs through the manipulation of precursor pathways, for instance via targeted chemical interactions with PNCs though addition of counter-ions. We are convinced that our approach holds potential to offer distinctive insights into the complex nucleation and growth mechanisms of related materials, encompassing MOFs, zeolites, covalent organic frameworks and myriad other materials. Also for proteins, where amorphous liquid-like precursors often precede supramolecular order, our approach is highly promising to disentangle the underlying mechanisms of crystallization and aggregation.

## MATERIALS AND METHODS

*Materials*

Zinc nitrate hexahydrate (98%, extra pure) was purchased from Acros Organics. 2-methylimidazole (99%) and Methanol (ACS spectrophotometric grade, >99.9%) were purchased from Sigma-Aldrich. Methanol was further purified through disposable syringe filters (Chromafil®, Glass fiber + Polyester filter, 0.20 μm pore size) purchased through Macherey-Nagel. The measurement cuvettes (High Precision Cell (QS), Light path 10x10mm) used for the *in situ* measurements were purchased through HellmaAnalytics.

Methanolic ZIF-8 synthesis was performed according to a published procedure.[6,7,11] While ZIF-8 was synthesized both *in situ* and *ex situ*, the procedure was similar. Two stock solutions were prepared, one containing the zinc source ($Zn(NO_3)_2 \cdot 6H_2O$) and another containing 2-methylimidazole ($C_4H_6N_2$). The two solutions were mixed to start crystallization. The ratio of addition for the synthesis was determined by the desired molar ratio in the form of 1:4:1000 for Zn:2-methylimidazole: methanol. For the *ex situ* synthesis the solution mixture was stirred for 24 hours at room temperature to allow nucleation and growth of the crystals. Afterwards, the crystals were separated from the reaction mixture by centrifugation and washed with methanol. This procedure was repeated 3 times followed by drying up under vacuum at 50°C for 24 hours. The *in situ* synthesis followed a similar procedure, however an automated syringe system was used to mix the two stock solutions with a 1:1 volume ratio and inject the mixture into the quartz measurement cell with a volume of 3 mL. The start of the mixing and subsequent injection into the measurement cell was used as the starting time of the reaction, but it took 5 seconds to complete injection and fill the measurement cell.

*HRS Theory*

SHS is a nonlinear optical process in which two photons of frequency ω interact simultaneously with a material to create a new photon at double the frequency 2ω. Analogously, for THS, a three photon interaction results in a photon at triple frequency 3ω. To overcome the low probability of both processes, SHS and THS typically require illumination with pulsed laser light of much higher intensity when compared to classical (linear) light scattering.

For nanoparticles that are small compared to the wavelength of the incident light, as is the case in this study (see ESI.1a), hyper-Rayleigh scattering ensues. In this regime scattering is isotropic, meaning that the light intensity is identical for all scattering angles in the case of vertically polarized incident light. The total scattering intensity is then found as the incoherent sum over all particles:

$$I_{\text{SHS}} \sim N < \left|\chi^{(2)}_{HRS}\right|^2 > \langle V^2 \rangle I_\omega^2 \qquad (M1)$$

$$I_{\text{THS}} \sim N < \left|\chi^{(3)}_{HRS}\right|^2 > \langle V^2 \rangle I_\omega^3 \qquad (M2)$$

$I_{\text{SHS}}$ and $I_{\text{THS}}$ are linearly related to the particle concentration N and quadratically to the orientationally averaged second and third susceptibility. $I_\omega$ is the incident light intensity. Within the electric dipole approximation, SHS is forbidden for structures with inversion symmetry, as the second susceptibility $\chi^{(2)} = 0$ in this case. In contrast to SHS, there are no symmetry restrictions for THS. For a single particle the intensity of both processes scales with the particle volume

(V) squared. For a collection of particles with a distribution of sizes, the intensity is therefore related to the average of the volume squared.

The scattering intensities for second- and third order processes are additive. As a result, all solution species will in principle contribute to the measured scattering intensity. Equations 3 and 4 describe the total SHS and THS intensity found for a generalized crystallization process:

$$I_{SHS} \sim (\sum_m N_m < |\beta_{HRS,m}^{\square}|^2 > + \sum_{ncs,i} N_i < |\chi_{HRS,i}^{(2)}|^2 > \langle V_i^2 \rangle) I_\omega^2 \quad (M3)$$

$$I_{THS} \sim (\sum_n N_n < |\gamma_{HRS,n}^{\square}|^2 > + \sum_{ncs,i} N_i < |\chi_{HRS,i}^{(3)}|^2 > \langle V_i^2 \rangle + \sum_{cs,j} N_j < |\chi_{HRS,j}^{(3)}|^2 > \langle V_j^2 \rangle) I_\omega^3 \quad (M4)$$

These equations encompass the response of all solution species (with $N_{m,n,i,j}$ described as number of solution species per unit volume). Molecular species (like solvent and linker) react as per $\beta$ and $\gamma$, their first and second hyperpolarizabilities. Macroscopic clusters and particles respond through their macroscopic susceptibilities $\chi^{(2)}$ and $\chi^{(3)}$. We have distinguished centrosymmetric (cs) and non-centrosymmetric (ncs) macroscopic contributions; the former contributing exclusively to THS, while the latter affects both SHS and THS. Notably, in crystallization, the macroscopic responses ($\chi^{(2)}$, $\chi^{(3)}$) often overshadow molecular ones ($\beta$, $\gamma$) early in the process, due to their volume-squared dependence.

Due to the tensorial nature of the first and second hyperpolarizability for molecules, and the second and third susceptibility for particles, polarized measurements contain additional information regarding the symmetry of species. In this study, the depolarization ratio $\rho$, defined as the ratio between the intensity of cross-polarized and parallel-polarized SHS and THS light, is used to interrogate symmetry.[33,35,36]

$$\rho_{SHS} = \frac{I_{HV}(2\omega)}{I_{VV}(2\omega)} \quad (M5)$$

$$\rho_{THS} = \frac{I_{HV}(3\omega)}{I_{VV}(3\omega)} \quad (M6)$$

In these equations the first subscript refers to the polarization of the scattered light, the second to the polarization of the incident light. H stands for horizontal polarization, V for vertical polarization. The SHS depolarization ratio ranges from 1/9 for dipolar to 2/3 for octupolar species.[40] For ZIF-8, which has octupolar $T_d$ symmetry (I-43m space group), the depolarization ratio assumes a value of 2/3. In THS, this ratio varies from 5/8 for hexadecapolar symmetry to 0 for isotropic species. For amorphous particles a depolarization ratio of exactly 0 is expected. For $T_d$ symmetry there is no exact solution. However, an almost isotropic response, with values near 0, can generally be assumed.[40]

### HRS and SLS setup

The experimental setup to perform the light scattering measurements was described elsewhere.[51] For our use, this setup was slightly modified.[38] This setup uses a femtosecond laser (Insight DS+, Spectra-Physics) with tunable wavelength ranging from 680 to 1300 nm, as a high-power light source. For the measurements, a fundamental wavelength of 1260 nm was used at a frequency of 80 MHz width pulse widths of 120 fs. The intensity variation of the incident light and the vertical polarization was accomplished by the combination of an achromatic half-wave plate and a Glan-Taylor polarizer. For the *in situ* measurements of ZIF-8, the intensity of the incident beam was kept constant at 1 W. The light is subsequently focused into a 10 x10 mm quartz cuvette containing the solution by an aspheric lens (A220TM, Thorlabs). The resulting beam waist of the focused beam is around 8 μm and a Rayleigh length of about 190 μm.

A secondary continuous wave laser at a wavelength of 543 nm was coupled into the system for static light scattering. Again, a half-wave plate and a Glan-Taylor polarizer combination was used to ensure a vertical polarization and power modulation of the incoming light. The power of this laser light was tuned to ensure that the EMCCD camera was not saturated throughout the crystallization reaction and was set at a constant power once the optimal intensity was achieved. The secondary beam was guided and aligned to ensure the same focus positions as the primary incoming beam. However, obtained focus size was different compared to the primary beam due to the absence of the confocal effect.

The scattered light was collected under 90° and subsequently collimated and focused onto the slit of a spectrograph (Bruker IS 500) attached to an EMCCD camera (Ixon 897, Andor Oxford instruments). To collect the full image of the focal point onto the slit, a rotation of the image by 90° was achieved by a set of mirrors in a periscope system. Before the spectrograph, a Wollaston prism was placed to separate the scattered light into its vertical and horizontal polarized components. Using the grating (50 l/mm, 300 blaze) in the spectrograph allowed to separate the incoming light, allowing us to measure simultaneously the depolarization ratio for SHS, THS and SLS. The temporal resolution of the in situ measurements were limited, due to the low scattering efficiency of SHS and THS. Therefore, the data acquisition times were always kept at 4s.

### Dynamic light scattering measurements

The time-resolved *in situ* measurements of methanolic ZIF-8 synthesis with dynamic light scattering was performed with a NanoBrook Omni Particle size and zeta potential analyzer (Brookhaven Instruments, Holtville NY, United States; 40 mW maximum power, 640 nm nominal wavelength red diode laser, measurement angle 173°). To achieve high temporal resolution for the *in situ* measurements, the correlograms were averaged for 10 seconds. Particle Solutions software was used to extract the number distributions and intensity of the measurements.

### Powder X-ray Diffraction

X-ray diffraction (XRD) patterns were obtained on a Malvern PANalytical Empyrean diffractometer, equipped with a PIXcel3D solid-state detector using a Cu anode (using Ka1 and Ka2). The powder samples were loaded onto a 48-well sample holder and X-ray diffractograms were recorded at room temperature in a transmission geometry within a 1.3 - 45° 2θ range using a step size of 0.013°.

*In situ NMR measurements*

*In situ* $^1$H NMR measurements were performed on a Bruker 800 MHz Avance Neo spectrometer equipped with a 5mm $^1$H/$^2$H/X BBO probehead ($^1$H Larmor frequency 801.25 MHz). Equal volumes of 2-methyl imidazole and zinc nitrate solutions in deuterated methanol (Methanol-D$_4$, Sigma Aldrich) were mixed in a 1:4 molar ratio in a 5 mm NMR tube (Norrel) and inserted into the probehead. $^1$H spectra were acquired every 5 s with an RF pulse of 18 kHz, an acquisition time of 1s and relaxation delay of 4s. The spectra were processed and integrated using Bruker Topspin 4.0.9 software.

*Ex-situ NMR measurements*

Concentrations of linker molecule (2-methylimidazole, 2-Me-ImH) were determined through $^1$H nuclear magnetic resonance (NMR) measurements using a Magritek SpinSolve 80 Carbon benchtop spectrometer ($^1$H basic frequency 80.478 MHz). Prior to the measurements of the ZIF-8 samples, a calibration curve was generated to determine the accuracy and precision of concentration determination through NMR. The internal standard employed for this purpose was tetramethylammonium bromide (TMAB, stock solution of 49.52 mM in D$_2$O). Acquisition parameters for all recorded datasets were: relaxation delay time = 15 s, pulse angle = 30°, acquisition time = 3.2 s, number of scans = 32, dummy scans = 0, total measurement time = 8 min. Processing of the NMR spectra were performed using Bruker's TopSpin (version 4.1.3) and involved zero-filling the free induction decay (FID) signal up to a total of 256k real datapoints (*si 256k*), exponential multiplication (lb 0.3 Hz), Fourier transform, automatic phase correction and baseline correction. Analysis of the spectra is described in the paragraph below.

Calibration curve

2-Me-ImH samples were weighed on an analytical balance (Mettler Toledo XS105). The weighted samples were dissolved in 650 μL deuterated solvent (600 μL D$_2$O, Sigma Aldrich, 99 atom %D, containing the TMAB + 50 μL D$_2$SO$_4$ solution, Sigma Aldrich 96-98 wt. % in D$_2$O, 99.5 atom %D). Clear solutions were obtained through shaking, vortex and sonification. The range of 2-Me-ImH concentrations for the calibration curve was [22.7 mM, 142.8 mM]. The second data point was replicated (n = 3), to allow estimation of weighing and operator error. The NMR tubes were filled with 500 μL of the solution to allow optimal shimming of the sample. Methyl signals of both TMAB and 2-Me-ImH were integrated in an unbiased way by the following two steps: 1. peak center and full-width at half maximum (FWHM) were determined with the *peakw* command, 2. FWHM was multiplied by nine and both added to and subtracted from the peak center defining the integration range. Simultaneously, the signal-to-noise ratio (SINO) of the 2-Me-ImH methyl signal was computed to ensure sufficient quality of the recorded spectra. Next, the integration values were normalized to their respective proton count (TMAB = 12 H's, 2-Me-ImH = 3 H's). Finally, the calculated ratios (2-Me-ImH/TMAB) between integration values of 2-Me-ImH and TMAB were compared with the expected ratios based on their relative concentrations.

Sample measurements

The ZIF-8 sample was weighed in quintuplicate followed by computation of the mean mass of the sample. Subsequently, the sample was dissolved in 650 μL deuterated solvent (600 μL D$_2$O containing the TMAB + 50 μL D$_2$SO$_4$). The presence of sulfuric acid disrupts the interaction between the Zn$^{2+}$ metal nodes and the linker molecule. Clear solutions were obtained through shaking, vortex and sonification. The same analysis protocol was followed for the calibration curve. The final calculated ratio was corrected using the calibration curve.

## ASSOCIATED CONTENT

**Supporting Information.** Additional information including detailed experimental and complementary results, theory supporting the methods and theoretical models can be found in the supporting information. This material is available free of charge via the Internet at http://pubs.acs.org.

## AUTHOR INFORMATION

Corresponding Author

* Correspondence and requests for materials should be addressed to S.V.C. (email: *Stijn.vancleuvenbergen@kuleuven.be*) or E.B. (email: *eric.breynaert@kuleuven.be*)

Author Contributions

All authors have given approval to the final version of the manuscript.

## ACKNOWLEDGMENT

C.K. acknowledges the Flemish Government for long-term Methusalem structural funding. E.B., and C.K. acknowledge joint funding by the Flemish Science Foundation (FWO; G083318N) and the Austrian Science Fund (FWF) (funder ID 10.13039/501100002428, project ZeoDirect I 3680-N34). E.B. acknowledges FWO for a "Krediet aan navorsers" 1.5.061.18N. NMRCoRe is supported by the Hercules Foundation (AKUL/13/21), by the Flemish Government as an international research infrastructure (I001321N), and by Department EWI via the Hermes Fund (AH.2016.134). The authors acknowledge FWO Vlaanderen and the teams at DUBBLE beamlines (ESRF, Grenoble) for financial and experimental support, respectively. Y.D.C. acknowledges the support from the Fund for Scientific Research-Flanders (FWO) (1234222N). S.V.C. thanks FWO Vlaanderen for financial support (12R8218N), and acknowledges KU Leuven for financing equipment required for harmonic light scattering (3E230460). A.D. acknowledges FWO Vlaanderen (G086522N) and KU Leuven (3E190382) for financial support.

## REFERENCES

(1) Moosavi, S. M.; Nandy, A.; Jablonka, K. M.; Ongari, D.; Janet, J. P.; Boyd, P. G.; Lee, Y.; Smit, B.; Kulik, H. J. Understanding the Diversity of the Metal-Organic Framework Ecosystem. *Nat. Commun.* **2020**, *11* (1), 1–10. https://doi.org/10.1038/s41467-020-17755-8.
(2) Eddaoudi, M.; Moler, D. B.; Li, H.; Chen, B.; Reineke, T. M.; O'Keeffe, M.; Yaghi, O. M. Modular Chemistry: Secondary Building Units as a Basis for the Design of Highly Porous and Robust Metal-Organic Carboxylate Frameworks. *Acc. Chem. Res.* **2001**, *34* (4), 319–330. https://doi.org/10.1021/ar000034b.

(3) Van Vleet, M. J.; Weng, T.; Li, X.; Schmidt, J. R. In Situ, Time-Resolved, and Mechanistic Studies of Metal–Organic Framework Nucleation and Growth. *Chem. Rev.* **2018**, *118*, 3681–3721. https://doi.org/10.1021/acs.chemrev.7b00582.

(4) Liu, X.; Chee, S. W.; Raj, S.; Sawczyk, M.; Král, P.; Mirsaidov, U. Three-Step Nucleation of Metal–Organic Framework Nanocrystals. *Proc. Natl. Acad. Sci. U. S. A.* **2021**, *118* (10), 1–7. https://doi.org/10.1073/pnas.2008880118.

(5) Deacon, A.; Briquet, L.; Malankowska, M.; Massingberd-Mundy, F.; Rudi, S.; Hyde, T.; Cavaye, H.; Coronas, J.; Poulston, S.; Johnson, T. Understanding the ZIF-L to ZIF-8 Transformation from Fundamentals to Fully Costed Kilogram-Scale Production. *Commun. Chem.* **2022**, *5* (18), 1–10. https://doi.org/https://doi.org/10.1038/s42004-021-00613-z.

(6) Cravillon, J.; Münzer, S.; Lohmeier, S. J.; Feldhoff, A.; Huber, K.; Wiebcke, M. Rapid Room-Temperature Synthesis and Characterization of Nanocrystals of a Prototypical Zeolitic Imidazolate Framework. *Chem. Mater.* **2009**, *21* (8), 1410–1412. https://doi.org/10.1021/cm900166h.

(7) Cravillon, J.; Schröder, C. A.; Nayuk, R.; Gummel, J.; Huber, K.; Wiebcke, M. Fast Nucleation and Growth of ZIF-8 Nanocrystals Monitored by Time-Resolved in Situ Small-Angle and Wide-Angle X-Ray Scattering. *Angew. Chem. Int. Ed. Engl.* **2011**, *50* (35), 8067–8071. https://doi.org/10.1002/anie.201102071.

(8) Cravillon, J.; Schröder, C. A.; Nayuk, R.; Gummel, J.; Huber, K.; Wiebcke, M. Fast Nucleation and Growth of ZIF-8 Nanocrystals Monitored by Time-Resolved in Situ Small-Angle and Wide-Angle X-Ray Scattering. *Angew. Chemie - Int. Ed.* **2011**, *50* (35), 8067–8071. https://doi.org/10.1002/anie.201102071.

(9) Cravillon, J.; Schröder, C. A.; Bux, H.; Rothkirch, A.; Caro, J.; Wiebcke, M. Formate Modulated Solvothermal Synthesis of ZIF-8 Investigated Using Time-Resolved in Situ X-Ray Diffraction and Scanning Electron Microscopy. *CrystEngComm* **2012**, *14* (2), 492–498. https://doi.org/10.1039/c1ce06002c.

(10) Terban, M. W.; Banerjee, D.; Ghose, S.; Medasani, B.; Shukla, A.; Legg, B. A.; Zhou, Y.; Zhu, Z.; Sushko, M. L.; De Yoreo, J. J.; Liu, J.; Thallapally, P. K.; Billinge, S. J. L. Early Stage Structural Development of Prototypical Zeolitic Imidazolate Framework (ZIF) in Solution. *Nanoscale* **2018**, *10* (9), 4291–4300. https://doi.org/10.1039/c7nr07949d.

(11) Cravillon, J.; Nayuk, R.; Springer, S.; Feldhoff, A.; Huber, K.; Wiebcke, M. Controlling Zeolitic Imidazolate Framework Nano- and Microcrystal Formation: Insight into Crystal Growth by Time-Resolved in Situ Static Light Scattering. *Chem. Mater.* **2011**, *23* (8), 2130–2141. https://doi.org/10.1021/cm103571y.

(12) Carraro, F.; Williams, J. D.; Linares-Moreau, M.; Parise, C.; Liang, W.; Amenitsch, H.; Doonan, C.; Kappe, C. O.; Falcaro, P. Continuous-Flow Synthesis of ZIF-8 Biocomposites with Tunable Particle Size. *Angew. Chemie* **2020**, *132* (21), 8200–8204. https://doi.org/10.1002/ange.202000678.

(13) Bustamante, E. L.; Fernández, J. L.; Zamaro, J. M. Influence of the Solvent in the Synthesis of Zeolitic Imidazolate Framework-8 (ZIF-8) Nanocrystals at Room Temperature. *J. Colloid Interface Sci.* **2014**, *424*, 37–43. https://doi.org/10.1016/j.jcis.2014.03.014.

(14) Wu, X.; Yue, H.; Zhang, Y.; Gao, X.; Li, X.; Wang, L.; Cao, Y.; Hou, M.; An, H.; Zhang, L.; Li, S.; Ma, J.; Lin, H.; Fu, Y.; Gu, H.; Lou, W.; Wei, W.; Zare, R. N.; Ge, J. Packaging and Delivering Enzymes by Amorphous Metal-Organic Frameworks. *Nat. Commun.* **2019**, *10* (1), 1–8. https://doi.org/10.1038/s41467-019-13153-x.

(15) Lim, I. H.; Schrader, W.; Schüth, F. Insights into the Molecular Assembly of Zeolitic Imidazolate Frameworks by ESI-MS. *Chem. Mater.* **2015**, *27*, 3088–3095. https://doi.org/10.1021/acs.chemmater.5b00614.

(16) Zhu, Y.; Ciston, J.; Zheng, B.; Miao, X.; Czarnik, C.; Pan, Y.; Sougrat, R. Unravelling Surface and Interfacial Structures of a Metal – Organic Framework by Transmission Electron Microscopy. **2017**, *16* (February), 4–9. https://doi.org/10.1038/NMAT4852.

(17) Patterson, J. P.; Abellan, P.; Denny, M. S.; Park, C.; Browning, N. D.; Cohen, S. M.; Evans, J. E.; Gianneschi, N. C. Observing the Growth of Metal – Organic Frameworks by in Situ Liquid Cell Transmission Electron Microscopy. *J. Am. Chem. Soc.* **2015**, *137*, 7322–7328. https://doi.org/10.1021/jacs.5b00817.

(18) Jin, B.; Wang, S.; Boglaienko, D.; Zhang, Z.; Zhao, Q.; Ma, X.; Zhang, X.; Yoreo, J. J. De. The Role of Amorphous ZIF in ZIF-8 Crystallization Kinetics and Morphology. *J. Cryst. Growth* **2023**, *603* (July 2022), 126989. https://doi.org/10.1016/j.jcrysgro.2022.126989.

(19) Saha, S.; Springer, S.; Schweinefuß, M. E.; Pontoni, D.; Wiebcke, M.; Huber, K. Insight into Fast Nucleation and Growth of Zeolitic Imidazolate Framework-71 by in Situ Time-Resolved Light and X-Ray Scattering Experiments. *Cryst. Growth Des.* **2016**, *16* (4), 2002–2010. https://doi.org/10.1021/acs.cgd.5b01594.

(20) Saha, S.; Wiebcke, M.; Huber, K. Insight into Fast Nucleation and Growth of Zeolitic Imidazolate Framework-71 by in Situ Static Light Scattering at Variable Temperature and Kinetic Modeling. *Cryst. Growth Des.* **2018**, *18* (8), 4653–4661. https://doi.org/10.1021/acs.cgd.8b00626.

(21) Ogata, A. F.; Rakowski, A. M.; Carpenter, B. P.; Fishman, D. A.; Merham, J. G.; Hurst, P. J.; Patterson, J. P. Direct Observation of Amorphous Precursor Phases in the Nucleation of Protein-Metal-Organic Frameworks. *J. Am. Chem. Soc.* **2020**, *142* (3), 1433–1442. https://doi.org/10.1021/jacs.9b11371.

(22) Kumar, M.; Li, R.; Rimer, J. D. Assembly and Evolution of Amorphous Precursors in Zeolite L Crystallization. *Chem. Mater.* **2016**, *28* (6), 1714–1727. https://doi.org/10.1021/acs.chemmater.5b04569.

(23) Rimer, J. D.; Tsapatsis, M. Nucleation of Open Framework Materials: Navigating the Voids. *MRS Bull.* **2016**, *41* (5), 393–398. https://doi.org/10.1557/mrs.2016.89.

(24) Jain, R.; Mallette, A. J.; Rimer, J. D. Controlling Nucleation Pathways in Zeolite Crystallization: Seeding Conceptual Methodologies for Advanced Materials Design. *J. Am. Chem. Soc.* **2021**, *143* (51), 21446–21460. https://doi.org/10.1021/jacs.1c11014.

(25) Olafson, K. N.; Li, R.; Alamani, B. G.; Rimer, J. D. Engineering Crystal Modifiers: Bridging Classical and Nonclassical Crystallization. *Chem. Mater.* **2016**, *28* (23), 8453–8465. https://doi.org/10.1021/acs.chemmater.6b03550.

(26) Rimer, J. D.; Chawla, A.; Le, T. T. Crystal Engineering for Catalysis. *Annu. Rev. Chem. Biomol. Eng.* **2018**, *9*, 283–309. https://doi.org/10.1146/annurev-chembioeng-060817-083953.

(27) Filez, M.; Caratelli, C.; Rivera-torrente, M.; Heck, A. J. R.; Van, V.; Bert, M.; Filez, M.; Caratelli, C.; Rivera-torrente, M.; Muniz-miranda, F.; Hoek, M.; Altelaar, M.; Heck, A. J. R.; Speybroeck, V. Van; Weckhuysen, B. M. Elucidation of the Pre-Nucleation Phase Directing Metal-Organic Framework Formation Elucidation of the Pre-Nucleation Phase Directing Metal-Organic Framework Formation. *Cell Reports Phys. Sci.* **2021**, *2* (12), 100680. https://doi.org/10.1016/j.xcrp.2021.100680.

(28) Öztürk, Z.; Filez, M.; Weckhuysen, B. M. Decoding Nucleation and Growth of Zeolitic Imidazolate Framework Thin Films with Atomic Force Microscopy and Vibrational Spectroscopy. *Chem. - A Eur. J.* **2017**, *23* (45), 10915–10924. https://doi.org/10.1002/chem.201702130.

(29) Yeung, H. H. M.; Sapnik, A. F.; Massingberd-Mundy, F.; Gaultois, M. W.; Wu, Y.; Fraser, D. A. X.; Henke, S.; Pallach, R.; Heidenreich, N.; Magdysyuk, O. V.; Vo, N. T.; Goodwin, A. L. Control of Metal–Organic Framework Crystallization by Metastable Intermediate Pre-Equilibrium Species. *Angew. Chemie - Int. Ed.* **2019**, *58* (2), 566–571. https://doi.org/10.1002/anie.201810039.

(30) Van Cleuvenbergen, S.; Smith, Z. J.; Deschaume, O.; Bartic, C.; Wachsmann-Hogiu, S.; Verbiest, T.; van der Veen, M. A. Morphology and Structure of ZIF-8 during Crystallisation Measured by Dynamic Angle-Resolved Second Harmonic

Scattering. *Nat. Commun.* **2018**, *9* (1), 1–9. https://doi.org/10.1038/s41467-018-05713-4.
(31) Van Cleuvenbergen, S.; Stassen, I.; Gobechiya, E.; Zhang, Y.; Markey, K.; De Vos, D. E.; Kirschhock, C.; Champagne, B.; Verbiest, T.; Van Der Veen, M. A. ZIF-8 as Nonlinear Optical Material: Influence of Structure and Synthesis. *Chem. Mater.* **2016**, *28* (9), 3203–3209. https://doi.org/10.1021/acs.chemmater.6b01087.
(32) Markey, K.; Putzeys, T.; Horcajada, P.; Devic, T.; Guillou, N.; Wübbenhorst, M.; Cleuvenbergen, S. Van; Verbiest, T.; De Vos, D. E.; Van Der Veen, M. A. Second Harmonic Generation Microscopy Reveals Hidden Polar Organization in Fl Uoride Doped MIL-53(Fe). *Dalt. Trans.* **2016**, *45*, 4401–4406. https://doi.org/10.1039/c5dt04632g.
(33) Dok, A. R.; Legat, T.; de Coene, Y.; van der Veen, M. A.; Verbiest, T.; Van Cleuvenbergen, S. Nonlinear Optical Probes of Nucleation and Crystal Growth: Recent Progress and Future Prospects. *J. Mater. Chem. C* **2021**, *9* (35), 11553–11568. https://doi.org/10.1039/d1tc02007b.
(34) De Coene, Y.; Deschaume, O.; Jooken, S.; Seré, S.; Van Cleuvenbergen, S.; Bartic, C.; Verbiest, T.; Clays, K. Advent of Plasmonic Behavior: Dynamically Tracking the Formation of Gold Nanoparticles through Nonlinear Spectroscopy. *Chem. Mater.* **2020**, *32* (17), 7327–7337. https://doi.org/10.1021/acs.chemmater.0c02178.
(35) Moris, M.; Van Den Eede, M. P.; Koeckelberghs, G.; Deschaume, O.; Bartic, C.; Van Cleuvenbergen, S.; Clays, K.; Verbiest, T. Harmonic Light Scattering Study Reveals Structured Clusters upon the Supramolecular Aggregation of Regioregular Poly(3-Alkylthiophene). *Commun. Chem.* **2019**, *2* (1), 1–9. https://doi.org/10.1038/s42004-019-0230-4.
(36) Moris, M.; Van Den Eede, M. P.; Koeckelberghs, G.; Deschaume, O.; Bartic, C.; Clays, K.; Van Cleuvenbergen, S.; Verbiest, T. Unraveling the Supramolecular Organization Mechanism of Chiral Star-Shaped Poly(3-Alkylthiophene). *Macromolecules* **2020**, *53* (21), 9513–9520. https://doi.org/10.1021/acs.macromol.0c01232.
(37) Moris, M.; Van Den Eede, M. P.; Koeckelberghs, G.; Deschaume, O.; Bartic, C.; Clays, K.; Van Cleuvenbergen, S.; Verbiest, T. Solvent Role in the Self-Assembly of Poly(3-Alkylthiophene): A Harmonic Light Scattering Study. *Macromolecules* **2021**, *54* (5), 2477–2484. https://doi.org/10.1021/acs.macromol.0c02544.
(38) Aerts, K.; Deschaume, O.; Bartic, C.; van Cleuvenbergen, S.; Clays, K.; Koeckelberghs, G.; Verbiest, T.; de Coene, Y. Multimodal Optical Analysis of Regioregular Poly(3-Hexylthiophene)s Reveals Peculiar Aggregation Dynamics. *Macromolecules* **2024**, *57* (4), 1521–1531. https://doi.org/10.1021/acs.macromol.3c02320.
(39) Van Steerteghem, N.; Clays, K.; Verbiest, T.; Van Cleuvenbergen, S. Third-Harmonic Scattering for Fast and Sensitive Screening of the Second Hyperpolarizability in Solution. *Anal. Chem.* **2017**, *89* (5), 2964–2971. https://doi.org/10.1021/acs.analchem.6b04429.
(40) Thierry Verbiest, Koen Clays, V. R. *Second-Order Nonlinear Optical Characterization Techniques Le*; CRC Press, 2009. https://doi.org/https://doi.org/10.1201/9781420070736.
(41) Harris, K. D. M.; Hughes, C. E.; Williams, P. A.; Gregory, R. Nmr Crystallography ' NMR Crystallization ': In-Situ NMR Techniques for Time-Resolved Monitoring of Crystallization Processes Nmr Crystallography. **2017**, 137–148. https://doi.org/10.1107/S2053229616019811.
(42) Hughes, C. E.; Williams, P. A.; Kariuki, B. M.; Harris, K. D. M. Establishing the Transitory Existence of Amorphous Phases in Crystallization Pathways by the CLASSIC NMR Technique **. **2018**, 1–6. https://doi.org/10.1002/cphc.201800976.
(43) Pellens, N.; Doppelhammer, N.; Radhakrishnan, S.; Asselman, K.; Chandran, C. V.; Vandenabeele, D.; Jakoby, B.; Martens, J. A.; Taulelle, F.; Reichel, E. K.; Breynaert, E.; Kirschhock, C. E. A. Nucleation of Porous Crystals from Ion-Paired Prenucleation Clusters. *Chem. Mater.* **2022**, *34*, 7139–7149. https://doi.org/10.1021/acs.chemmater.2c00418.
(44) Wang, F.; Richards, V. N.; Shields, S. P.; Buhro, W. E. Kinetics and Mechanisms of Aggregative Nanocrystal Growth. *Chem. Mater.* **2014**, *26* (1), 5–21. https://doi.org/10.1021/CM402139R/ASSET/IMAGES/LARGE/CM-2013-02139R_0017.JPEG.
(45) Yeung, H. H. -M.; Sapnik, A. F.; Massingberd-Mundy, F.; Gaultois, M. W.; Wu, Y.; Fraser, D. A. X.; Henke, S.; Pallach, R.; Heidenreich, N.; Magdysyuk, O. V.; Vo, N. T.; Goodwin, A. L. Control of Metal–Organic Framework Crystallization by Metastable Intermediate Pre-equilibrium Species. *Angew. Chemie - Int. Ed.* **2019**, *58*, 566–571. https://doi.org/10.1002/ange.201810039.
(46) Houlleberghs, M.; Hoffmann, A.; Dom, D.; Kirschhock, C. E. A.; Taulelle, F.; Martens, J. A.; Breynaert, E. Absolute Quantification of Water in Microporous Solids with 1H Magic Angle Spinning NMR and Standard Addition. *Anal. Chem.* **2017**, *89* (13), 6940–6943. https://doi.org/10.1021/acs.analchem.7b01653.
(47) Vanderschaeghe, H.; Houlleberghs, M.; Verheyden, L.; Dom, D.; Chandran, C. V.; Radhakrishnan, S.; Martens, J. A.; Breynaert, E. Absolute Quantification of Residual Solvent in Mesoporous Silica Drug Formulations Using Magic-Angle Spinning NMR Spectroscopy. *Anal. Chem.* **2023**, *95*, 1880–1887. https://doi.org/10.1021/acs.analchem.2c03646.
(48) Grunewald, C.; Seidel, R.; Chapartegui-arias, A. Observation of Early ZIF-8 Crystallization Stages with X-Ray Absorption Spectroscopy †. **2020**, 8–11. https://doi.org/10.1039/d0sm01356k.
(49) Venna, S. R.; Jasinski, J. B.; Carreon, M. A. Structural Evolution of Zeolitic Imidazolate Framework-8. *J. Am. Chem. Soc.* **2010**, *132* (51), 18030–18033. https://doi.org/10.1021/ja109268m.
(50) Priandani, L.; Aliefa, A.; Arjasa, O. P.; Pambudi, F. I. An Experimental and Computational Study of Zeolitic Imidazolate Framework (ZIF-8) Synthesis Modulated with Sodium Chloride and Its Interaction with $CO_2$. *Bull. Chem. React. Eng. Catal.* **2023**, *18* (3), 521–538. https://doi.org/10.9767/bcrec.20033.
(51) Coene, Y. De; Cleuvenbergen, S. Van; Steerteghem, N. Van; Baekelandt, V.; Verbiest, T.; Bartic, C.; Clays, K. Fluorescence-Free Spectral Dispersion of the Molecular First Hyperpolarizability of Bacteriorhodopsin. *J. Phys. Chem. C* **2017**, *121*, 6909–6915. https://doi.org/10.1021/acs.jpcc.7b00625.

Insert Table of Contents artwork here

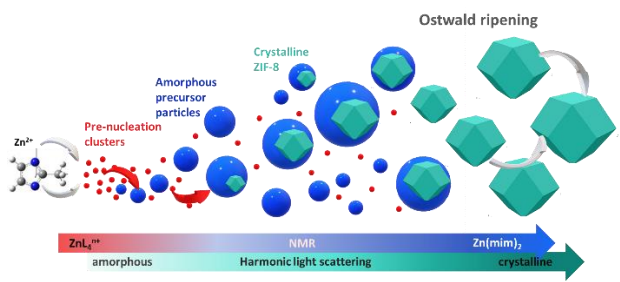